\documentclass[]{tMPH2e}

 \usepackage{graphics}

 \usepackage{graphicx}

\usepackage{amssymb}

\usepackage{color}

\begin{document}

\title{Study of dipole moments of LiSr and KRb molecules by quantum Monte Carlo methods}

\author{Shi Guo$^{a}$  
, Michal Bajdich $^{b}$  Lubos Mitas $^{a}$  and Peter J. Reynolds$^{a c}$ $^{\ast}$  \thanks{$^\ast$Corresponding author. Email: peter.j.reynolds16.civ@mail.mil}\\
$^{a}$ {\em{ Department of Physics, North Carolina State University, Raleigh, NC 27695}};
$^{b}${\em{Joint Center for Artificial Photosynthesis, Lawrence Berkeley Laboratory, and Department of Chemical and Biomolecular Engineering, University of California, Berkeley, CA 94720}}; 
$^{c}${\em{Physics Division and Physical Sciences Directorate, Army Research Office, Research Triangle Park, NC 27703}}
\\\vspace{6pt}\received{final version received 15 March 2013} }

\maketitle

\begin{abstract}
Heteronuclear dimers are of significant interest to experiments seeking to exploit ultracold polar molecules
in a number of novel ways including precision measurement, quantum computing, and quantum simulation. 
We calculate highly accurate Born-Oppenheimer total energies and electric dipole moments 
as a function of internuclear separation for two such dimers, LiSr and KRb.  
We apply fully-correlated, high-accuracy
quantum Monte Carlo methods for evaluating these molecular properties in a many-body framework.
We use small-core effective potentials combined with multi-reference Slater-Jastrow trial 
wave functions to provide accurate nodes for the fixed-node diffusion Monte Carlo method. 
For reference and comparison, we calculate the same properties with Hartree-Fock and 
with restricted Configuration Interaction methods, and carefully assess the impact of the 
recovered many-body correlations on the calculated quantities.  For LiSr we find a highly nonlinear dipole moment curve, which may make this molecule's
dipole moment tunable through  vibrational state control.

\end{abstract}


\section{Introduction}
\label{}

 Motivated both by the desire for deeper understanding of basic physics and of numerous potential applications,
there has been a great deal of work over the last few years devoted to cooling small molecules
 \cite{Carr} and molecular ions \cite{ERH}.
This follows up on a few decades of exciting progress in cooling and trapping of atoms and atomic ions.  
Molecules, of course, are far more challenging, due to their complex internal structure, particularly their many internal degrees of freedom, such as vibrational and 
rotational levels, in addition to fine and hyperfine structure.   A number of approaches have been pursued, with varying degrees of success.  While making molecules
 cold has been achieved in numerous ways, reaching the ultra-cold regime, and particularly the quantum degenerate regime, has been limited.  

The most general methods of cooling molecules use buffer gases and supersonic expansion \cite{WJD,JMD,JD}, or velocity filtering \cite{RSA}
techniques.
Another direct approach is Stark decceleration \cite{BHL}.  On the other hand, direct laser cooling of molecules, in analogy to the very successful approach for neutral 
atoms, was long deemed impractical due to the complex level structures. The simplest and most widespread method of direct laser cooling of atoms is Doppler cooling, where
radiative forces originate from momentum transfer to atoms from a laser field, and subsequent spontaneous emission of slightly higher energy photons into random directions.  
Repeating this optical cycle tens of thousands of times cools neutral atoms very quickly to the Doppler limit (which is mass dependent, but typically reaches sub-mK temperatures). 
For molecules, the problem with the conventional scheme is that excited states can radiatively decay out to a multitude of other states.  
This leads to decays that destroy any closed cycling transitions.  
Exciting population from all these states back to the starting point requires an
 impractically large number of lasers.  Only recently has laser cooling 
of a special class of molecules been 
demonstrated \cite{Shuman,Shuman_2} based on insights 
from earlier work \cite{MDDR,BKS}. 

Most successful at reaching into the ultra-cold regime are methods 
that create the molecules from previously cooled atoms. However, 
the number of molecules produced in this way is fairly 
small $(\approx 10^4)$.  
Such methods include photoassociation \cite{JMS} and magnetoassociation (exploiting Feshbach resonances) \cite{HJM}. 
 While the former has been a well-established technique to produce homonuclear dimers from cold atoms, its application to produce heteronuclear molecules
 from two different species of laser-cooled alkali atoms is more recent.  Such diatomic molecules can be polar, and this is of particular interest
 (see, for example, reviews in \cite{Carr, JMD, Dulieu,Dulieu_2}).  

The usefulness of polar molecules arises because they, unlike neutral atoms, interact 
via the characteristic long-range, anisotropic dipolar interaction, 
making them controllable by external electric fields.  
On the other hand, they are less strongly coupled to the environment than ions, 
making them less prone to decoherence \cite{Pupillo}. This makes 
them attractive for quantum manipulation, such as for quantum information
 processing \cite{14,15,16, 35,Rabl2006,Rabl2006_2}, and precision 
measurements to test symmetries \cite{17,18,18_1,18_2} 
and the constancy of fundamental ``constants" \cite{19,20,21,22,23,alpha}.  

Additional interest lies in alkali-alkali earth dimers such as LiSr, 
one of the molecules studied here. In contrast with alkali dimers, 
the unpaired spin makes them able to be manipulated with both external electric and magnetic fields.  
This provides an avenue for different physics to be explored, whether in fundamental tests,  as qubits, 
in optical lattices where the competing interactions are important, or potentially even for use in atomic clocks 
where atoms such as Sr are already showing great potential \cite{clocks}.  

Cooling, trapping and quantum manipulation of polar molecules will have an important impact on a diverse range of fields beyond just fundamental tests and quantum
 computing.  One such example is in condensed matter physics, through quantum ``emulation" or simulation of many-body quantum physics that is inaccessible to
 even today's (and any day's) high-performance computers, let alone to analytical solution (for example \cite{OLE,25,28}). Cold polar molecules trapped in optical
 lattices make an extremely interesting many-body system (with long-range dipole-dipole interactions) promising a rich variety of novel quantum phases, 
such as supersolids and topologically ordered states \cite{25,24}.  One can imagine many other novel strongly-correlated quantum phases, particularly in reduced 
dimensions \cite{10,10_1,10_2}.  Local control of the density or orientation of polar molecules may allow one to create or simulate charge density waves or (pseudo-) spin-density waves, or even a random, glassy system.

Another and rapidly emerging area is ultra-cold chemistry \cite{HJM,Dulieu,26,27,39}.  In the regime where the de Broglie wavelength of the molecule becomes comparable, 
or even orders of magnitude larger than the molecule itself, the classical notion of a reaction coordinate is at odds with quantum mechanics.   Coherent population transfer 
methods (e.g., STIRAP) have already produced a gas of fermionic 
$^{40}$K$^{87}$Rb molecules with a temperature of a few hundred nano-Kelvin at a phase-space density 
getting very close to that 
needed to achieve degeneracy.  In this regime one expects 
entirely new phenomena in looking at chemical reactions at ultralow energies \cite{9}. In addition to just potential energy surfaces being relevant, the ultracold
 reaction rate is controlled by quantum statistics and quantum coherence effects, including tunneling.  Moreover, quantum statistics only allows certain
 collisions, and tunneling leads to threshold laws.  In this regime, quantum control also can take on new meaning, with resonance-mediated reactions and 
collective many-body effects becoming prominent.  Polar molecules also provide a handle for control through application of relatively weak electric fields. 

Among applications, cold molecules hold great promise for improving sensors of all sorts.  So far, all demonstrated matter-wave interferometric sensors utilize atoms.
 Using dipolar molecules instead of atoms could lead to orders-of-magnitude improvements in the sensitivity of such sensors \cite{2}. The advantage comes from
 the ability to guide the molecules with modest electric field gradients, thereby creating steeper confining potentials. Steeper potential gradients allow one to
 load waveguide structures more efficiently; this leads to a larger number of particles and better statistical sensitivity.  Also because of the steeper potentials,
 molecules can be kept further away from the wires and guiding surfaces, thus improving performance, as these are major sources of decoherence.  Moreover, because
 sensitivity (e.g., to rotation) is proportional to the area enclosed by an interferometer, a large-area storage ring (with electrostatic guiding) \cite{3} 
becomes a possible interferometer. 

Bi-alkali molecules produced thus far include RbCs \cite{JMS,Kerman2004}, KRb \cite{7,31,31_2,36}, NaCs \cite{32} and LiCs \cite{33,Kraft2006}.  
Photoassociation and magnetoassociation can be combined with STIRAP to allow coherent population transfer into a low-energy bound state, and significantly
 enhances the rate of molecular production \cite{7,36,37,38}.   

The strength of the dipole moment in these diatomics is critical to their possible uses, ranging from their potential as qubits, their use in precision measurement experiments,
the nature of their quantum phase transitions, as well as simply their ability to be manipulated. 

This motivates our study, which is focused on applying first principles 
methods to two polar diatomics, one a bi-alkali, 
and one an alkali-alkali earth.  Specifically, we use quantum Monte Carlo in combination with basis set methods
to look carefully at KRb and LiSr, both in their ground states ($X^1\Sigma^+$ and $X^2\Sigma^+$, respectively).

\section{Methods}

Since the bonds in all these systems are weak, and consist of a mixture of covalent and 
van der Waals bonding mechanisms, first principles calculations tend to be quite involved 
and highly sensitive to basis sets, degree of correlation included, nature of approximations, and other factors. Density Functional Theory (DFT) 
for van der Waals systems is often less than satisfactory,
typically showing a varying degree of overbinding without any obvious systematic trends. Several new
DFT functionals with corrections for dispersion interactions have been proposed
very recently; see for example, \cite{Scheffler, Lundqvist}. However, thorough 
testing and benchmarking of these new DFT approaches is necessary before 
they can be reliably applied across a variety of systems. Dispersion interactions
result from transient-induced polarizations between the interacting constituents, and 
are therefore subtle many-body effects which are difficult to capture in the functionals
framework. 

On the other hand, the powerful arsenal of basis-set correlated methods, such as Configuration 
Interaction (CI), have their own challenges for systems such as these.  
For example, the dipole moment can be very sensitive to the basis set, and convergence for weakly bonded 
systems can be very slow \cite{Halkier}.  In addition, the dipole moment appears highly sensitive 
to the level of correlation used, especially for problems which require multi-reference 
treatment.  This is well known, but a systematic study can be found in Ref. \cite{Halkier}.  Our CI results here also demonstrate this.  It is therefore important to explore other types of methods to understand
the impact of many-body effects more thoroughly.  A highly accurate 
alternative approach is quantum Monte Carlo (QMC) \cite{Reynolds1982, Reynolds_book}. 
QMC is very attractive since it is in principle exact; in practice due to approximations it has a residual weak sensitivity to the size of basis sets, and it
captures the correlations at a level of 90-95\% \cite{Reynolds1982, WF, MB}.  That is something that is quite difficult to achieve 
by correlated methods based on expansions in basis sets. 

In fact, there are previous studies of 
molecular dipole moments calculated by QMC methods for a few molecular systems.   
Schautz and colleagues carried out QMC study of the carbon monoxide molecule and obtained
 a very good estimate for this non-trivial problem in which correlation reverses the sign 
of the dipole obtained at the Hartree-Fock level \cite{FS}. The dipole moment of the lithium hydride
molecule was 
 computed by fixed-node diffusion Monte Carlo (DMC) in a good agreement with experiment \cite{SL}. 
Recently, several transition metal 
monoxide molecules have also been calculated by QMC methods. Besides the binding energies 
and equilibrium bond lengths, the dipole moments were also calculated, all with reasonably good agreement 
to experiment \cite{LucasWagner}. In addition, transition dipole moments such as for the Li atom
were calculated by QMC approaches some time ago \cite{Barnett92}. 

\subsection{Sampling}

For our calculations here, we employ the two most common QMC methods, 
variational and fixed-node diffusion Monte Carlo
(VMC and DMC) \cite{Reynolds1982, Reynolds_book, Reynolds_VMC_review, WF, MB, Needs, Kolorenc}.  We summarize just the main points of the algorithms.
VMC is based on the variational principle, and the Monte Carlo method of
evaluation of multi-variate integrals.  Given a trial wavefunction $\Psi_T(R)$, we sample
 a set of points in the 3$N$ dimensional space of electron ``configurations" according to the
probability density
$|\Psi_T(R)|^2$, where $R$ denotes the set of spatial coordinates of all electrons.
The expectation value of an operator $A$ is evaluated over $M$ samples, and is given by

\begin{displaymath}
\langle A \rangle_{VMC}=\frac{1}{M} \sum_{m=1}^M \Psi_T^{-1}(R_m)[A\Psi_T(R_m)] + \epsilon_{stat}
\end{displaymath}
where $m$ labels the samples (``random walkers"),
 and $ \epsilon_{stat}$ is the statistical
error, which scales as $O(1/\sqrt{M})$.\\

Diffusion Monte Carlo in contrast is  based on a stochastic process which solves
 the imaginary-time many-body Schrodinger equation, projecting out the ground state through iteration of a short-time Green's function as
\begin{displaymath}
f(R,t+\tau)=\int G(R \leftarrow R', \tau) f(R', t) dR',
\end{displaymath}
where $f(R,t)=\Phi(R,t)\Psi_T(R)$.  The wave function $\Phi(R,t\rightarrow\infty)$ is the solution we are seeking, while $\Psi_T(R)$ is a trial
function, 
and $G(R \leftarrow R', \tau)$ is the Green's function or propagator that evolves $f$ 
 \cite{Reynolds_book}. The fixed-node approximation is imposed
by preventing walkers from crossing the
nodal hypersurface of $\Psi_T$ (i.e., the zero locus of $\Psi_T(R)$). This condition restricts
$f(R,t)\ge 0$  so that the sampled probability density remains non-negative everywhere. In this 
manner we can evade the well-known fermion sign problem \cite{Reynolds1982}. 
Since the nodes of the trial function are not necessarily precisely those of the exact wave function, the  fixed-node condition
introduces a bias. From a large body of work using this method, we know that the typical size of the fixed-node bias
is between 5 and 10\% of the correlation energy, essentially for all systems, including
molecules, clusters and solids \cite{WF}.

\subsection{Trial functions}

The quality of the trial wave function plays two roles.  First, highly accurate trial functions lead to smaller
statistical fluctutations and faster QMC sampling and convergence.  Second, better trial functions typically go hand-in-hand 
with improvements of the nodal hypersurfaces, and hence result in smaller fixed-node bias.  
Clearly,
we want to use the best trial function we can. Traditional methods, including correlated approaches,
provide a convenient 
starting point in this respect.  In addition, however, quantum Monte Carlo 
can employ explicitly correlated wave functions as well, without prohibitive computational cost.
This enables us to 
reach beyond the limits of traditional approaches, even before beginning to project out the true (fixed node) ground state.

In our calculations, 
we have used the Slater-Jastrow trial wave functions
\begin{displaymath}
\Psi_T(R)=
\Psi_A(R)
e^{J(R)} 
\end{displaymath}
which embody both the traditional computational chemistry starting point and explicit correlation in the two factors.
The antisymmetric part $\Psi_A(R)$ is often given by a single configuration state function which is a
product of Slater spin-up and spin-down determinants, or, more generally,
by a linear combination of such configurations.
This linear combination of configurations  
with, say, single and double excitations, 
becomes
the Configuration Interaction (CI) expansion 
\begin{displaymath} 
\Psi_A=c_0\Psi_0+\sum_{ar} c_a^r \Psi_a^r+\sum_{a<b,r<s} c_{ab}^{rs}\Psi_{ab}^{rs}  .
\end{displaymath}
The orbitals for our Slater determinats 
are obtained from Hartree-Fock calculations, and 
our CI expansion is calculated using the GAMESS package. 

The Jastrow factor contains electron-ion, electron-electron,
and electron-electron-ion correlation terms.
The electron-electron correlation functions build in the correct electron-electron cusps.
The actual forms we use for the Jastrow factor 
are described in detail in Ref \cite{MB}. 
The Jastrow variational parameters and the CI expansion coefficients 
are optimized in VMC by minimizing a linear combination
of the variational energy and the variance of the local energy, $[H\Psi_T]/\Psi_T$. 
 The optimized trial function is then employed in our fixed-node 
DMC runs.  

\subsection{Mixed and pure expectation values}

In diffusion Monte Carlo, expectation values of quantities which commute with the Hamiltonian are exact \cite{Reynolds1982};
however, expectation values of non-commuting operators, $\hat{O}$, are estimated  
by mixed estimators, i.e., 
with the distribution $\Psi_T\Phi$, instead of $\Phi^2$. One can correct for this in a number of 
ways \cite{Barnett92, Reynolds_properties}.  A commonly used approximation is 
\begin{displaymath}
< \Phi| \hat{O} | \Phi> \approx 2< \Phi | \hat{O} | \Psi_T> - < \Psi_T| \hat{O} | \Psi_T>.
\end{displaymath}
Exact sampling of $\Phi^2$, while computationally more complex and significantly more costly, 
is also possible \cite{Kalos,WF,Reynolds_properties}. Accuracy gained 
from exact sampling methods is limited
by any approximations involved, both fundamental and technical. These include, e.g., the fixed-node 
restriction and the localization approximation in treating the nonlocal 
effective core potentials (see below).   Since the optimization of the wave function 
is done stochastically on finite samples, it is difficult to eliminate a possible optimization 
bias reflecting differences in the localization error with different Jastrows.  This is especially so in the case of weakly bonded systems such as LiSr (with bonding of the order 
of 0.006 a.u.).
Therefore, with any likely improvements to be invisible, the results below are based on the mixed estimator, which provides a good balance
between accuracy and efficiency of the calculations.    

\subsection{Effective core potentials}

For heavy atoms we use effective core potentials (ECPs) to eliminate the core electrons. In particular, 
we employ energy consistent, small-core ECPs \cite{MB1,MB2} for the atoms beyond the first row.   
For this work we explored ECPs with different core sizes, and
we found that the results appeared to converge with the small-core ECPs.   
In fact, two different sets of small-core ECPs provided
essentially the same overall picture of the electronic structure as described below.
In the context of the QMC calculations,  
the ECPs are localized by a projection onto the trial function, making the ECPs contributions 
to the local energy formally similar to the other terms in the Hamiltonian, as described in detail
elsewhere \cite{Reynolds87,Mitas91}. 

\section{Calculational Details}

The pseudopotential we ultimately used for the bulk of 
the calculations here removes small cores for the elements beyond the first row,
with the resulting configurations of valence electrons in KRb given as 
K($3s^23p^64s^1$) and Rb($4s^24p^65s^1$).  Gaussian basis sets used in the calculations
were gradually increased in size until we observed saturation to about $\approx$ 0.001 $E_h$
at the self-consistent level;
basis sets of 25s22p13d/[4s4p3d] were used for both K and Rb.
In the LiSr calculation, we kept all Li atom electrons in the valence space, 
while for Sr we eliminated 18 core electrons, 
so that the valence configuration became Sr$(4s^24p^65s^2)$. Quite extensive basis sets  
for Sr and Li, with sizes 7s6p5d1f/[7s4p2d1f] and 27s7p1d/[7s5p1d], provided 
saturated HF energies at the level of 0.001 $E_h$ or better.

A proper description of the dipole moments requires high-quality correlated methods.
This is true in particular for  
LiSr, which does not exhibit any binding at the  
HF level. 
(KRb, on the other hand, has a single $\sigma$ bond, so that using a single-reference wave function, and
CI with single and double excitations, was deemed to be appropriate in this case.)
For LiSr we explored several correlation
levels and sizes of active occupied and virtual spaces. In particular,
we carried out singles and doubles (SD) with
5 active occupied and 70 virtual orbitals; SD and triples (SDT) 
with 5 active occupied and 55 virtual orbitals;  SDT and quadruples (SDTQ) 
with 3 active occupied and 48 virtual orbitals.
Excitations beyond doubles proved to be important for the LiSr dipole moment.  Even with quadruples included,
it remained difficult to be certain that the correlations were adequately described.

This is where QMC methods provide an important alternative, and a powerful option
to probe for correlations beyond what is
practical with CI. 
For QMC, the CI wave functions only serve as a starting point---as trial functions.  
They are truncated to only their most significant terms by imposing 
a cutoff on 
the weights of the configuration state functions. 
Different cutoff values from 0.05 to 0.01 were tested. 
It turned out that the optimal cutoff value is around 0.03-0.05.  (Values smaller than this no longer decreased the QMC energy within the detectability of the error bars.)  Using
a cutoff of 0.05 for LiSr and 0.03 for KRb, we were able to improve the nodal
surfaces (evidenced by lower QMC energy) with reasonable efficiency and statistical consistency.

\section{Results and discussion}

The energy that we obtain as a function of bond length is shown in Fig. 1, for LiSr in Fig. 1a, and KRb in Fig. 1b.  We compare our best CI calculation with the diffusion QMC (DMC) results.  For LiSr, we show the QMC results with two different trial functions, DMC1 and DMC2.   
We see that the QMC correlation energy for both molecules is considerably better than that found by the CI method, and provides significant
overall improvement of the potential energy surfaces.  We show only correlated methods, as HF gives no binding at all for LiSr.

As previously discussed, the dipole moment of LiSr is particularly sensitive---to pretty much everything (basis set size, level of correlation in the theory, 
nodes in the QMC trial function, etc.).  We show in Fig. 2 this sensitivity with respect to amount of correlation in the theory. 
 As is clear, CI SDT, which is often sufficient, has not converged here.  
It would be difficult to know if even CI SDTQ is sufficient were it not for our QMC calculation.  
In Fig. 3a we compare the CI SDTQ result for the dipole moment with our two QMC calculations with different trial functions.
  While the two QMC results are not identical, they vary immensely less than the different CI calculations. 
 Moreover, the QMC with refined nodes from the multi-reference trial function gives almost the same results as CI SDTQ. 
 This provides additional confidence in this result.  This can be seen better in the expanded Figure 4,
 where we compare the dipole moment of LiSr as computed from only correlated methods, 
and compare our results against a recent result in the literature as well.

\begin{figure}

\begin{tabular}{cr}
\includegraphics[scale=0.29, ]{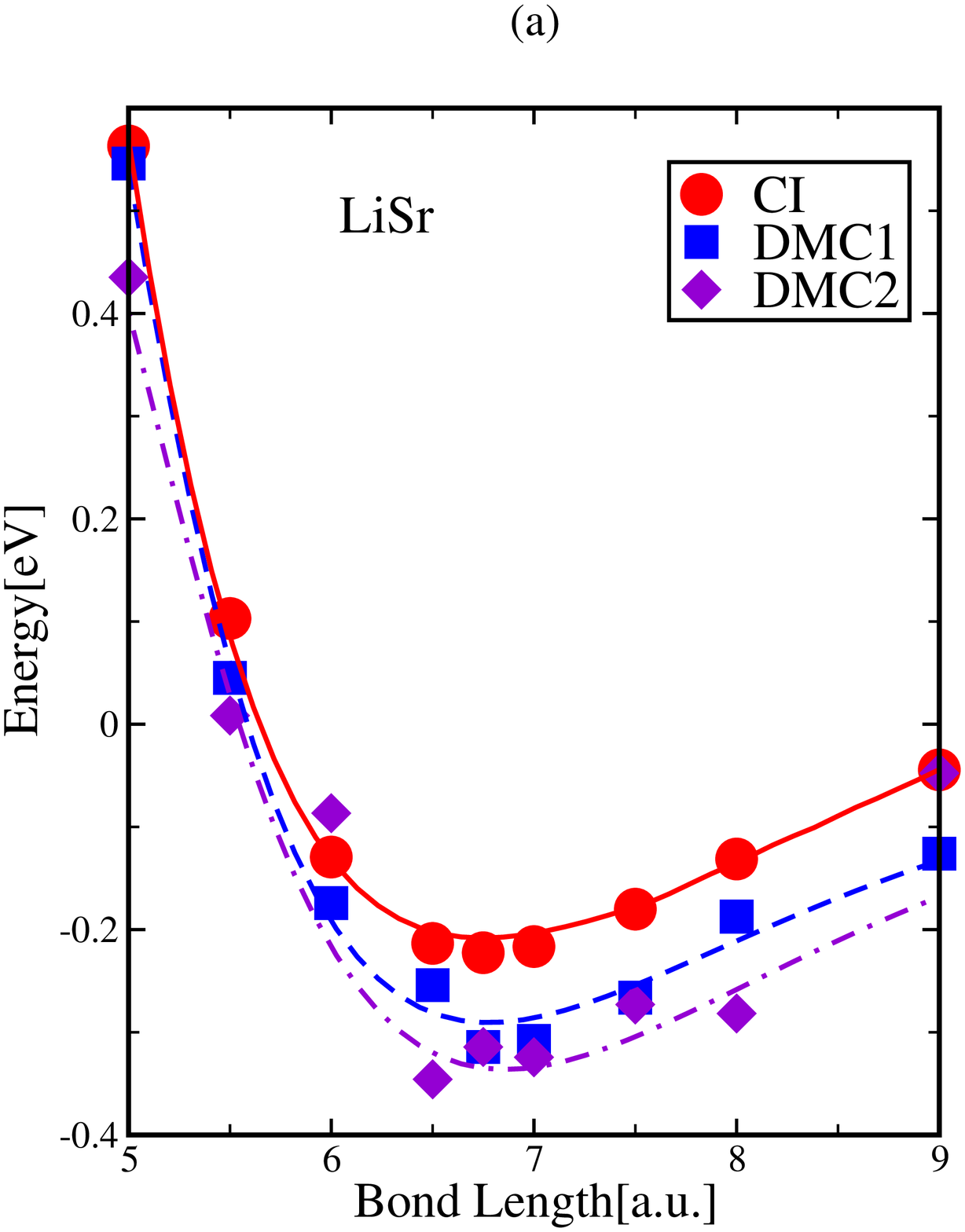} &
\includegraphics[scale=0.29, ]{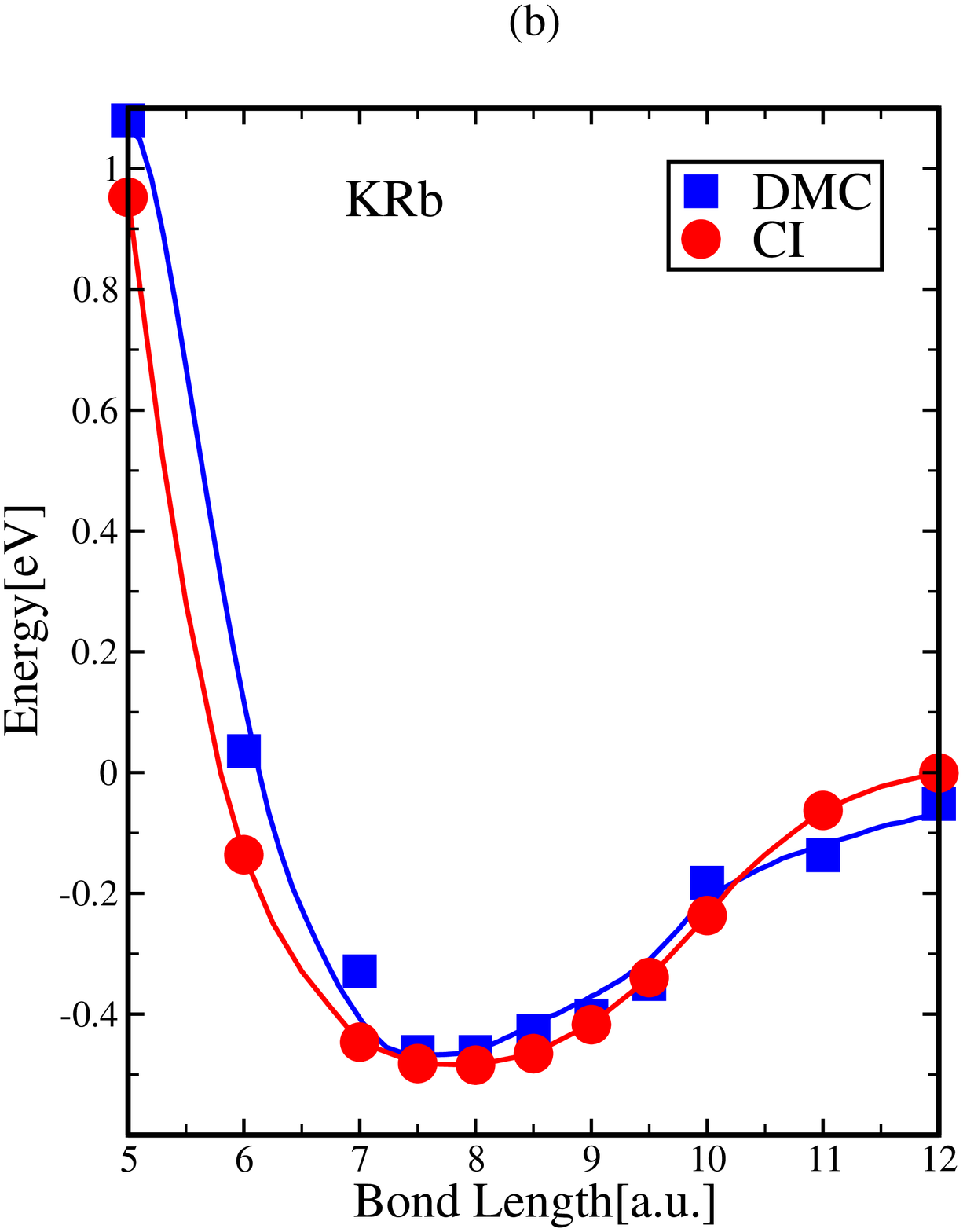}\\
\end{tabular}
\vspace{5mm}
\caption{ 
The binding energies of (a) LiSr and (b) KRb as a function of the bond length 
with different methods.
Two optimized Jastrow DMC trial wave functions are used for LiSr. DMC1 denotes a single Slater
determinant, while DMC2 is multi-reference with a
cutoff weight of 0.05. 
For KRb, the DMC trial wave function
is multi-reference, with a cutoff weight of 0.03, and again an optimized Jastrow. 
Here and in subsequent figures, the statistical error bars of the QMC results are approximately of the symbol 
size.  Additional deviations of the QMC results reflect the small errors discussed in Section 2.
The CI curves correspond to CI/SDTQ for LiSr and CI/SD for KRb.
} 
\end{figure}

\begin{figure}
\begin{center}
\includegraphics[scale=0.39, ]{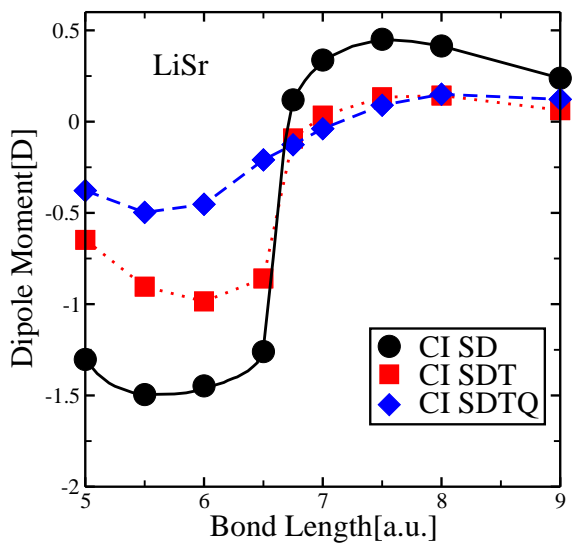}
\vspace{5mm}
\caption{
The dipole moment of LiSr calculated by the CI method with SD, SDT
and SDTQ levels of correlation, as described in the text.}
\end{center}
\end{figure}

\begin{figure}
\begin{tabular}{cr}
\includegraphics[scale=0.29, ]{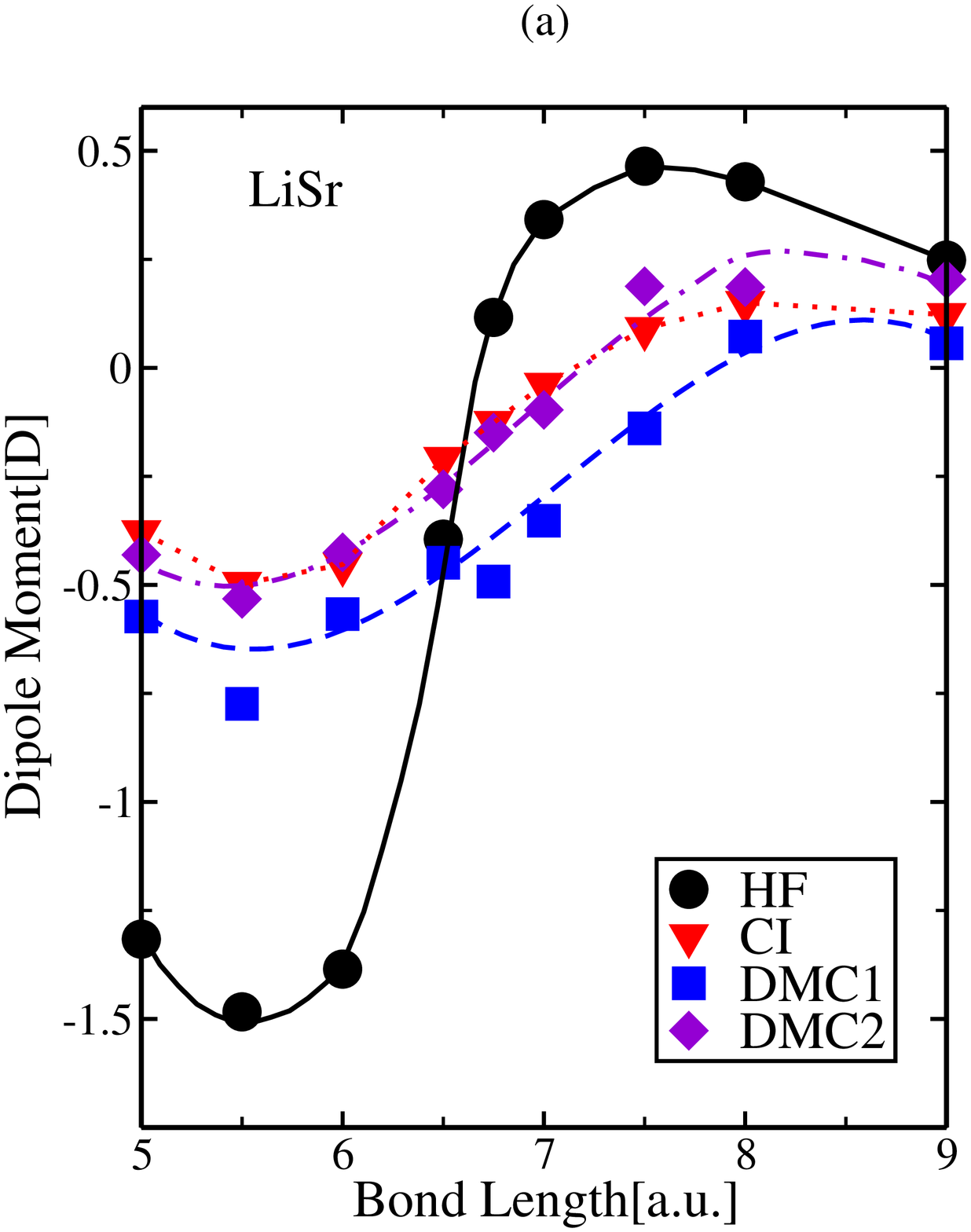} &
\includegraphics[scale=0.29, ]{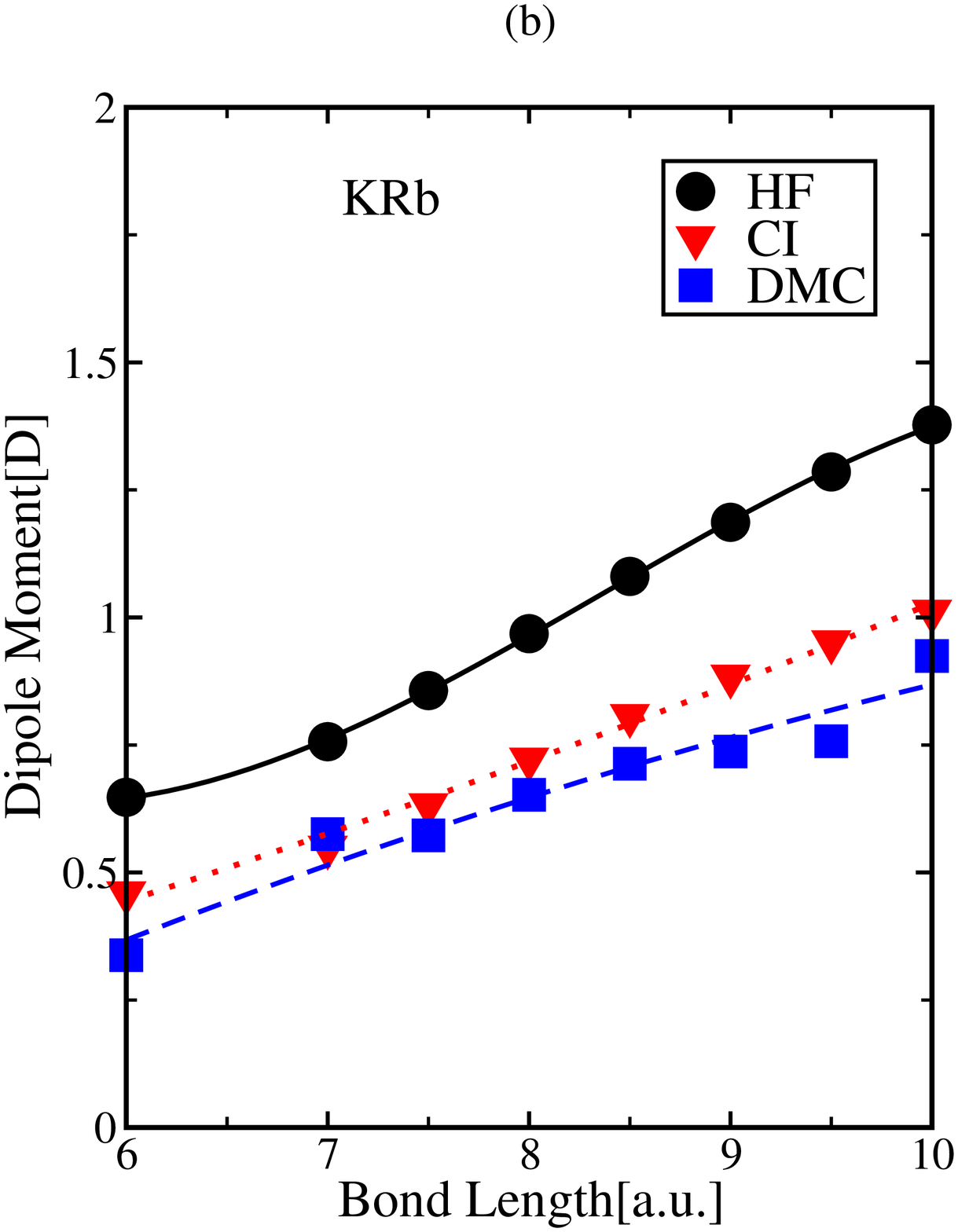} \\
\end{tabular}
\vspace{5mm}
\caption{ The dipole moment of (a) LiSr and (b) KRb as a function of 
internuclear distance, as obtained in DMC,  HF and CI 
methods (CI/SDTQ for LiSr and CI/SD for KRb). 
The trial wave functions, DMC, DMC1 and DMC2 are
the same as described in Fig. 1.
}

\end{figure}

\begin{figure}
 \begin{center}
  \includegraphics[scale=0.39, ]{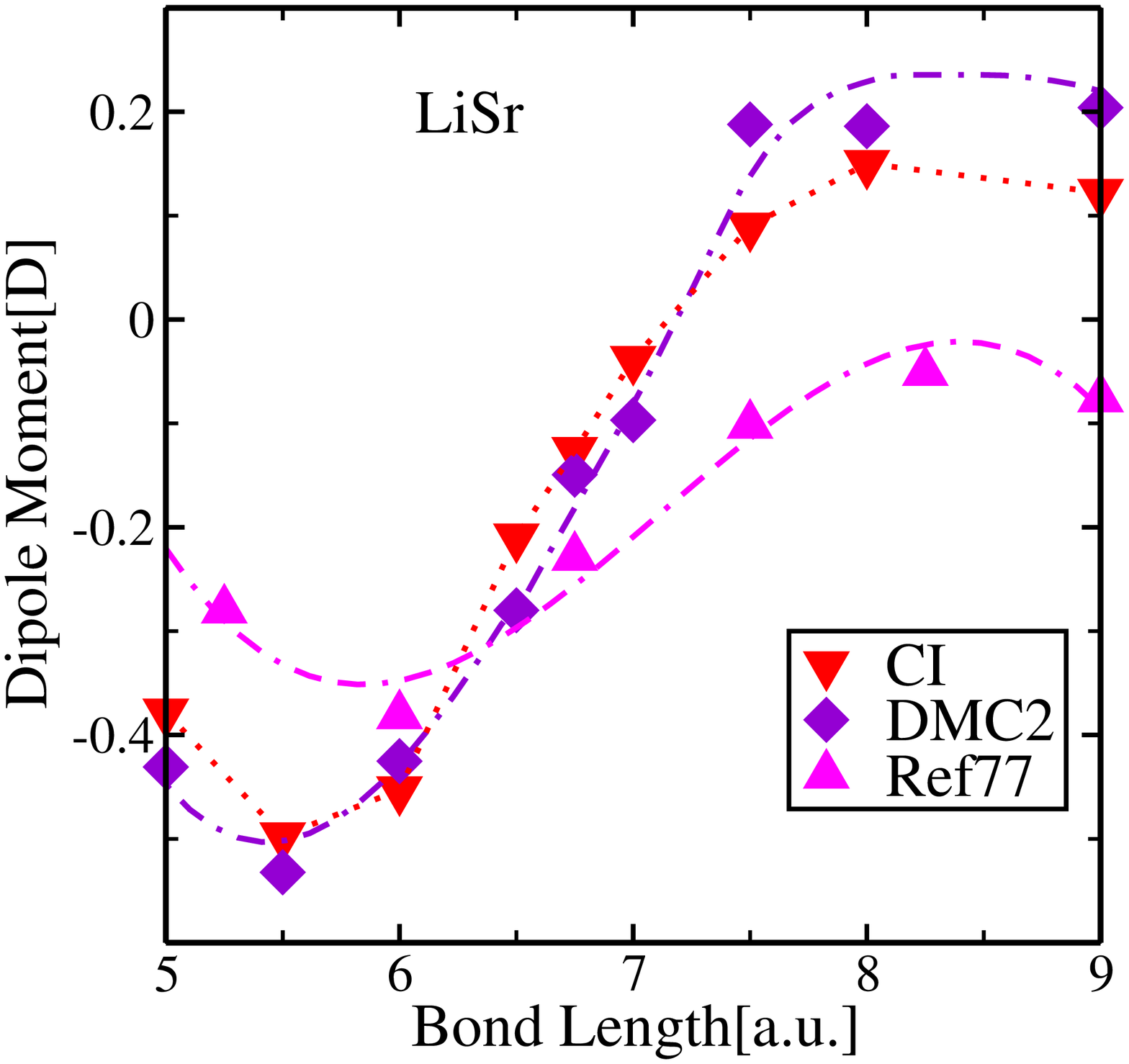}
\vspace{5mm}
\caption{ Dipole moment for LiSr from our best CI and QMC calculations compared with results 
from Ref.\cite{Kotochigova}.  The pronounced nonlinearity allows switching of the value of $d$ by state selection to a vibrational state near dissociation.  See text.}
 \end{center}

\end{figure}

Figure 3b shows the dipole moment results for KRb. 
The bonding behavior, as well as the dipole moment of KRb,
is rather straightforward to understand.  There is a single 
$\sigma$ bond formed from the K(4s) and Rb(5s) atomic states. 
The dipole moment monotonically increases with the bond length,
corresponding to growth of the distance between the effective charges,
and also some possible growing enhancement of the electron density 
in the region of the K atom.
This trend is observed already at the HF 
level, indicating that it is driven by the one-particle   
self-consistent balance of the energy contributions.
However, quantitatively, the HF dipole moment can be seen to be too large.  This is because
electron correlation
decreases the charge polarization over the whole range of calculated
interatomic distances. This can be equivalently 
understood as a result of the well-known HF bias towards larger
ionicity of the bonds. The absence of correlation tends to make the exchange more prominent, which drives the electronic structure 
towards stronger charge polarization. 
It is interesting to observe that the dipole moments computed from 
both of our correlated methods appear to be 
rather close, although the degree of description of the many-body
effects is very different, with the CI recovering only about 30\% of the correlation
energy compared to $\approx$  90\% in QMC.  

On the other hand, LiSr has a more complicated electronic 
structure due to the presence of the 
 two $(5s^2)$ electrons in the outermost shell of the Sr atom,
 with the resulting open-shell $X^2\Sigma^+$ molecular ground state. 
The last valence electron occupies an anti-bonding 
$\sigma$ level which is formed by the combination of 
Li(2s) and Sr(5s) atomic states. 
The overall bond is thus weaker, and this 
accounts for the lack of binding seen in
the HF method. Even after including $\approx 90$\%
of the correlation energy via QMC, as we have done here, the binding is small,
only of the order of 
$\approx$ 0.25 eV.  Even more interestingly, 
the dipole moment changes sign as a function of separation! This behavior
is visible in all three methods, implying that one can trace its origin to changes at the
single-particle level. Indeed, by plotting the 
isosurfaces of the Hartree-Fock 
HOMO (antibonding $\sigma$) and LUMO orbitals (see Figs. 5 and 6)  
we see that as the bond length increases, the nature of the orbitals change very significantly.
These changes lead to sizeable restructuring of the electron density 
already at the HF level, with the resulting sign change of the dipole moment as plotted in Figs. 2, 3a, and 4.

In addition, due to the reordering of the LUMO levels with varying bond length as 
illustrated in Fig.6, the CI expansion coefficients 
change significantly as well.
The leading excited configuration for bond length $d=5 $ a.u.  is the double excitation
$\sigma_{b}^2\sigma_{a} \to (\pi_{bx}^2 +\pi_{by}^2)\sigma_a$ where $\sigma_b$ and $\sigma_a$ denote the
highest bonding and anti-bonding valence molecular orbitals, respectively.  
One of the corresponding $\pi_b$ orbitals 
is plotted in the upper part of Fig. 6. 
On the other hand, 
for bond length of $d=7.5$ a.u., the lowest virtual state
is the next bonding orbital $2\sigma_b$, and the corresponding double excitation 
$\sigma_{b}^2\sigma_{a} \to 2\sigma_b^2\sigma_a$
becomes dominant. 
This restructuring, both in the one-particle (Fig. 6) and the many-body framework,
has a strong impact on the correlations. The dipole moment is affected, as is seen both
from smaller absolute values as compared to HF, as well as the shift of the sign change
towards longer bond lengths.
Overall, our results indicate that LiSr exhibits a sizeable dipole moment mainly for smaller interatomic 
distances, while at larger bond lengths the dipole decreases significantly.

\begin{figure}
\begin{center}
  \includegraphics[scale=0.12]{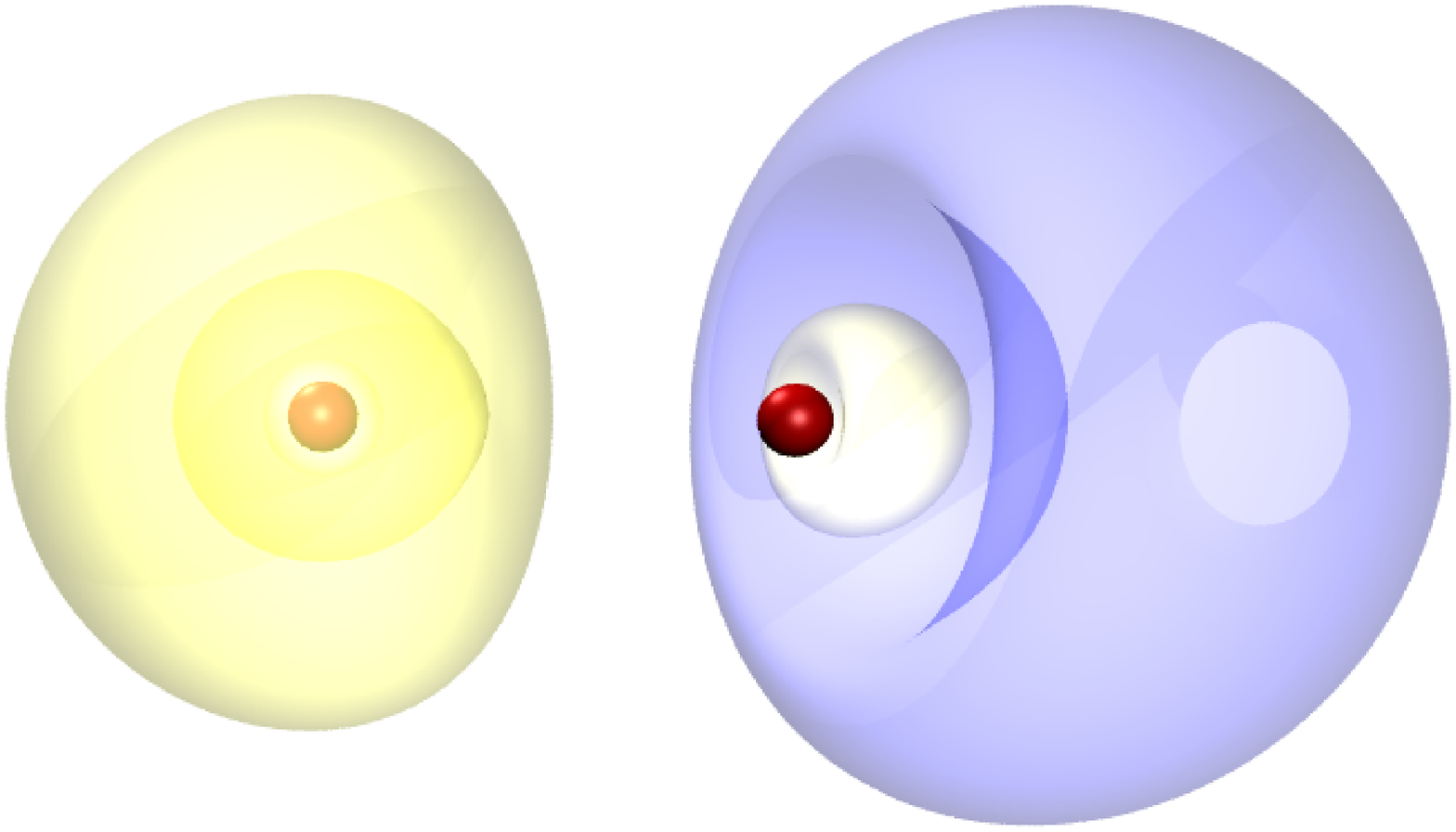}
 \hspace{5mm}
 \includegraphics[scale=0.12]{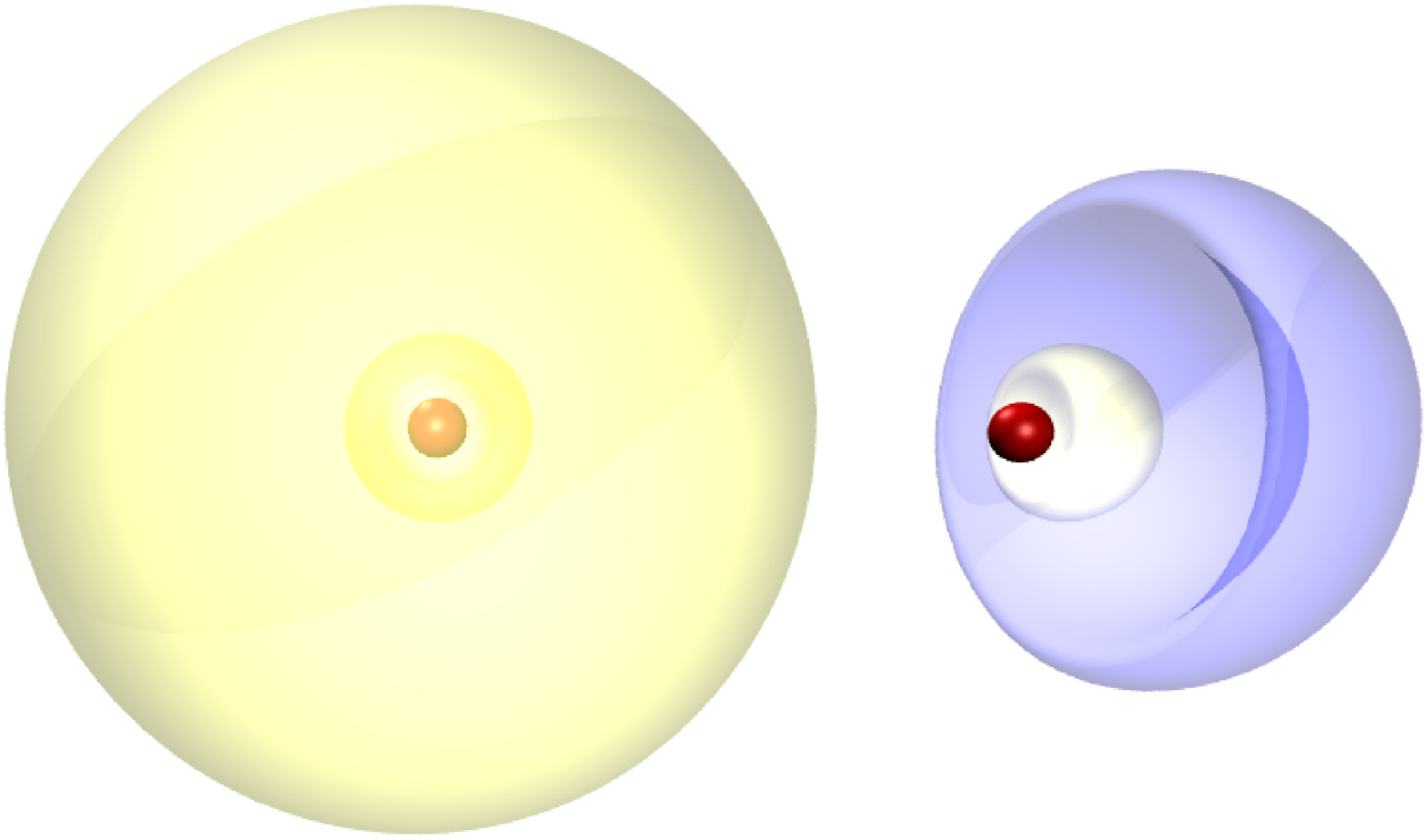} 
\caption{ Isosufaces of the positive and negative lobes of the highest occupied molecular orbital (HOMO)
of LiSr, for interatomic distances  5 a.u. (top) and 7.5 
a.u. (bottom). }
\end{center}
\end{figure}

\begin{figure}
\begin{center}
 \includegraphics[scale=0.12]{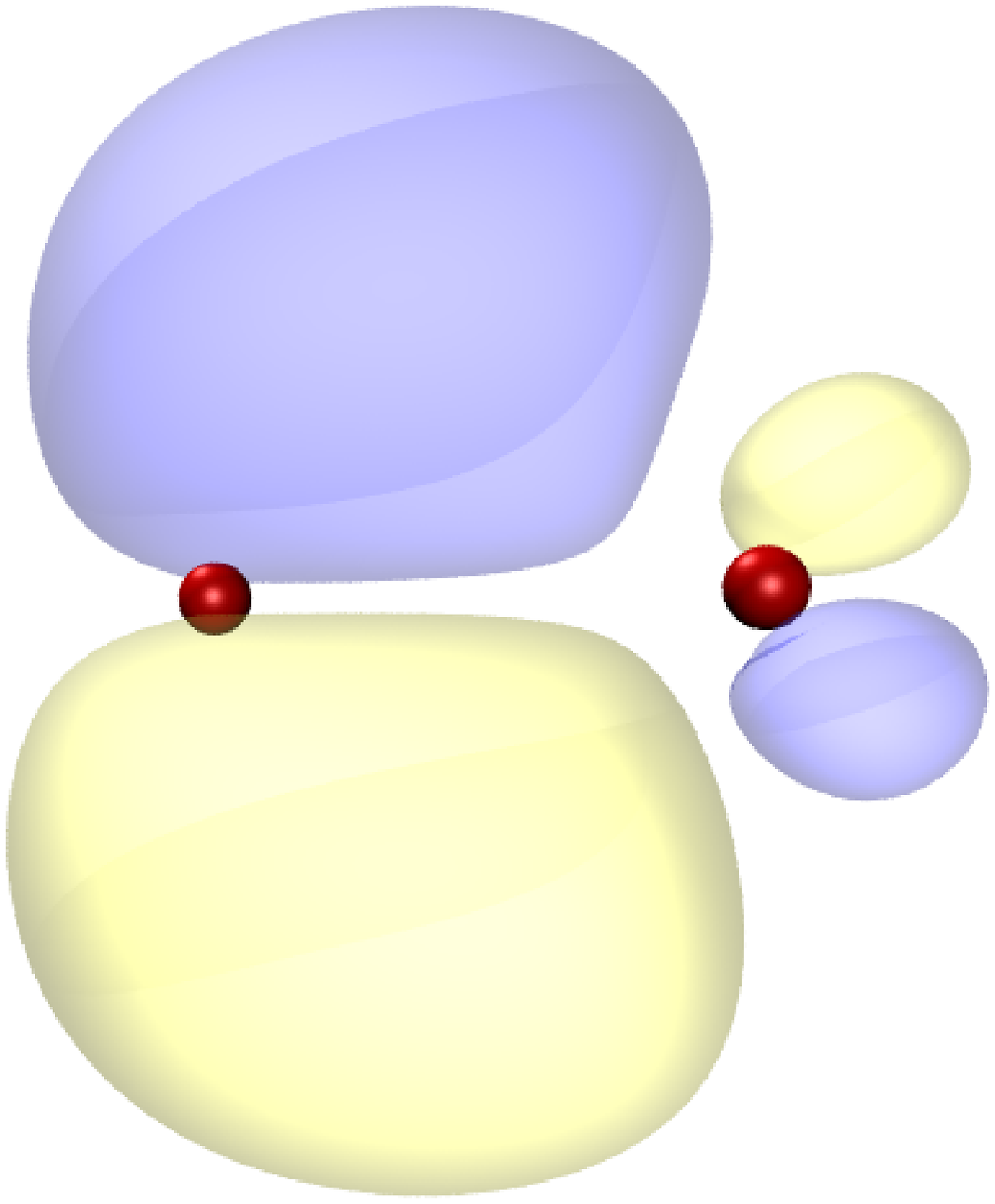}

\hspace{10mm}
 \includegraphics[scale=0.12]{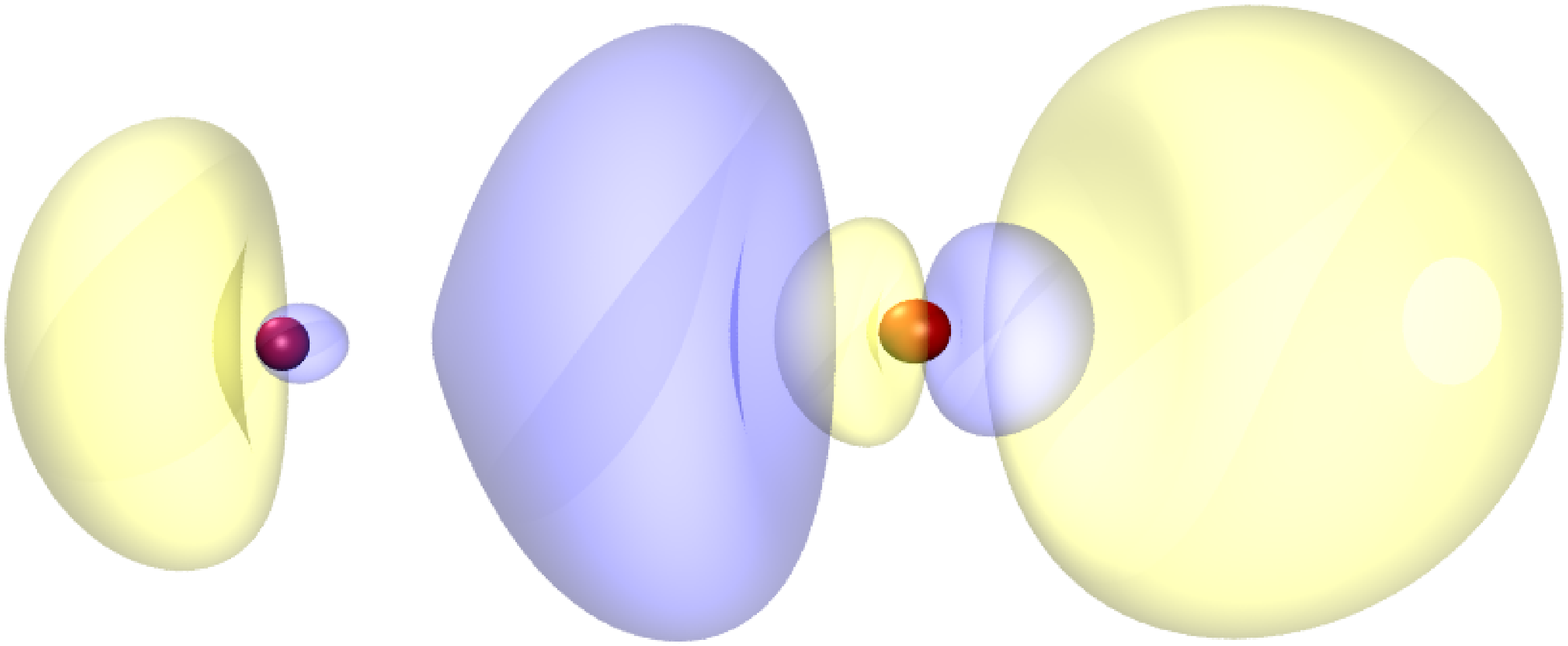}

\caption{ Isosurfaces of the positive and negative lobes of 
the lowest unoccupied molecular orbital (LUMO)
 of LiSr, for interatomic distances 5 a.u. (top) and 7.5 a.u. (bottom).}
\end{center}
\end{figure}

The quality of the trial wave function and its nodal surface has a significantly more pronounced 
impact on the LiSr dipole moment result than on the KRb moment, and one cannot rely on a single configuration 
for the former.  
The inclusion of the most important configurations from the
CI expansion improves the accuracy of the DMC estimations quite significantly.
It is interesting, if somewhat unexpected, that
our QMC results obtained with our best trial function, apart from the statistical noise, agree
remarkably well with our most accurate CI calculations. This consistency is reassuring since  
the employed methods are largely independent, 
and capture the electron correlations in very different manners. The need to employ  
triples and quadruples in CI indicates that correlations beyond  
 multi-reference configurations based on singles/doubles excitations
are important.
On the QMC side, the single configuration trial function already captures the main correction
to the HF dipole moment, due to the correlations, as can be seen in Fig 3a.  However, for high accuracy results,
the complicated orbital restructuring
of the LiSr molecule clearly requires a multi-reference treatment which includes the dominant 
non-dynamical effects explicitly, thereby improving the nodal surface. This accounts for the difference seen in Fig 3a between DMC1 and DMC2.
This observation is very much in line
with previous QMC calculations of other systems with significant multi-reference or near-degeneracy effects.

The spectroscopic constants for these molecules have been computed, and are 
shown in Tables 1 and 2, together with the most recently published calculations \cite{LiSr_calc2010,Kotochigova}
which used much less fully converged coupled cluster and CI methods.
For LiSr we calculated the dipole moment also in the first vibrationally excited state, and we found
that it is basically the same as in the ground state within the given error bar. This can be
understood from the fact that our results show the dipole being
approximately linear with bond length over the range $\approx$ 6.0 - 7.7 a.u.,
and this interval
covers the spatial range of both the ground and the first vibrational states.
Nevertheless, the nonlinearity of the LiSr dipole moment may be exploitable for vibrational dependent control of $d$.  
In particular, it appears that the highest bound vibrational states might have considerably 
smaller dipole moments, making 
it possible to largely switch off (and on) the dipole moment through tailored excitations.

The overall agreement between the calculational approaches,
considering how small the quantities are, is very reasonable.
The differences reflect the systematic biases of
the approaches used, and clearly illustrate
the challenge of describing weakly-bonded systems with high accuracy.  Since
these calculations reflect the state of the art of these computational methodologies, this shows
what is currently feasible. It is clearly desireable to increase the accuracy further, in order to
decrease the uncertainties. This can be accomplished with
further development of the methods; in particular, for QMC approaches one would need
much higher sensitivity in quantities which do not commute with the Hamiltonian so that the
optimization of much larger multi-reference wave functions would be feasible.
This is a promising direction for future explorations.

\begin{table}
\caption {Spectroscopic constants for the LiSr molecule:
bond length $R_e$, potential well depth $D_e$, harmonic constant $\omega_e$, and the ground state averaged
dipole moment $\langle d \rangle $.
The values in the row labeled DMC are calculated
from the DMC2 data displayed in Figs. 1a and 3a.  
Numbers in parentheses give the statistical uncertainty in the last significant figure.}
{
\vspace{3mm}
    \begin{tabular}{|c|c|c|c|c|}
        \hline
         ~   & $ R_e$ (a.u.) &  $D_e$ ($cm^{-1}$) & $\omega_e$($cm^{-1}$) &  $\langle d \rangle$ (D)  \\ \hline
        DMC  & 6.80(5)  & $2.7(3)\times10^3$  & 167(7) & - 0.14(2)  \\
        Ref \cite{Kotochigova} & 6.71  & $2.401\times10^3$ & 184 & - 0.244 \\
        Ref \cite{LiSr_calc2010} & 6.57  & $2.587\times10^3$  & 185 & - 0.340  \\
       \hline
    \end{tabular}
}
\end{table}

\begin{table}
{
\caption {Spectroscopic constants for the KRb molecule, defined as in Table 1. 
The values in the row labeled DMC correspond to the data displayed in
Figs. 1b and 3b. }
\vspace{3mm}
    \begin{tabular}{|c|c|c|c|c|}
        \hline
        ~           & $R_e$ (a.u.)&  $D_e$ ($cm^{-1}$) & $\omega_e$($cm^{-1}$) & $\langle d \rangle$(D) \\ \hline
        DMC         &  7.58(5) & $3.8(3)\times 10^3$ & 77(2)    &0.58(2) \\
        Ref \cite{LiSr_calc2010}        &  7.63 & $4.199\times 10^3$ & 76    &0.615 \\
        \hline
    \end{tabular}
}
\end{table}

\section{Conclusions.} 

We have carried out careful calculations of the dipole moments and potential energy curves for the KRb and LiSr molecules using Hartree-Fock, Configuration Interaction, and fixed-node
diffusion quantum Monte Carlo methods. 
The calculations show significant effects of the electronic correlations on the magnitude of
the dipole moments over the whole range of the investigated interatomic distances.

Although single determinant QMC already captures most of the correlation energy, 
the low-lying excitations need to be explicitly 
included into the trial function for the LiSr dimer, particularly for accurate computation of the dipole moment. 
The need is clear from the fact that the weights of some  
configurations beyond the reference HF configuration are significant.
Comparisons of CI(SD) 
with  CI(SDTQ)  clearly illustrates
the multi-reference nature of the ground state. 
The accuracy of our results is supported by the consistency
between our most extensively correlated CI approach and our methodologically quite distinct QMC results.

In addition, the high percentage of the correlation energy recovered by QMC ($\approx$ 90\%) captures
most of the dynamical correlation, and therefore provides much more extensive insight into 
how correlations affect sensitive molecular properties. 

{\bf Acknowledgments.} We gratefully acknowledge support by the U.S. Army Research Office.

\bibliographystyle{tMPH}

\end{document}